\documentclass[aps,prl,nofootinbib,floatfix,twocolumn,showpacs]{revtex4-1}

\usepackage{pstricks}
\usepackage{graphicx}
\usepackage{bm}
\usepackage{amsmath,amssymb,amsfonts}
\begin{document}

\title{Testing Planck-Scale Gravity with Accelerators}

\author{Vahagn Gharibyan}
\email[]{vahagn.gharibyan@desy.de}
\affiliation{Deutsches Elektronen-Synchrotron DESY - D-22603 Hamburg, Germany}

\begin{abstract}
 Quantum or torsion gravity models predict unusual properties of space-time at very short 
distances. In particular, near the Planck length, around $10^{-35}m$, empty space 
may behave as a crystal, singly or doubly refractive. However, this hypothesis 
remains uncheckable for any direct measurement since the smallest distance 
accessible in experiment is about $10^{-19}m$ at the LHC. Here I propose a 
laboratory test to measure the space refractivity and birefringence induced  
by gravity. A sensitivity from $10^{-31}m$ down to the Planck length 
could be reached at existent GeV and future TeV energy lepton accelerators 
using laser Compton scattering. 
There are already experimental hints for gravity signature at distances approaching 
the Planck length by 5--7 orders of magnitude, derived from SLC and HERA data. 
\end{abstract}

\maketitle

{\it Introduction.---}
The quantum formalism can not be directly applied to gravitation
and that is one of the major problems on a way of understanding and describing 
the physical reality.
An important reason for this is the dynamical space concept adopted in
general relativity, the currently accepted theory of gravity which states that any 
mass or particle modifies the space geometry (or metrics). On the other hand, the successful 
quantum theories within the Standard Model operate only in a fixed geometry space. 
For instance, observed violations of the discrete symmetries such as space, charge and 
time parities are attributed to the particles and their interactions while the scene 
of the interactions, the space-time, is considered to remain perfectly 
symmetric~\cite{Amsler:2008zzb}. These two faces of space 
are believed to unify at distances near the Planck length
$l_P =1.6\cdot 10^{-35}m$  (or mass $M_P = 1.2\cdot 10^{19}GeV$, natural units are 
assumed throughout the Letter). 
At this scale gravity is expected to be similar in strength to the
electroweak and strong forces and quantum effects become important for
the gravitational field. String theory and loop quantum gravity
theory are prominent candidates which set a framework to make
predictions in that energy domain. In many cases, unconventional
space-time properties are suggested, such as vacuum 
refractivity~\cite{AmelinoCamelia:1997gz} 
and/or birefringence~\cite{Gambini:1998it}. As suggested in this Letter, such effects
may be studied by
using high energy accelerator beams. The proposed experiments probe the vacuum symmetry 
in a search for a handedness or chirality of the empty space presumed by quantum gravity.
One particular example is circular birefringence \hbox{$\Delta n = n_L - n_R$} of 
space, with $n_{L(R)}$ being the refraction index of left(right) helicity photons 
traversing the space. 
In the following, quantitative theoretical estimates and existing experimental limits are 
quoted, the formalism of the proposed method is presented, and the sensitivity of 
accelerator experiments is discussed.

{\it Quantum and torsion gravity predictions.---}
Since Planck mass $M_P = \sqrt{c \hbar /G}$ is built from the speed of light and fundamental 
Planck and gravitational constants, this mass scale is considered to be relativistic 
and quantum gravitational. 
Most general modification of photon dispersion relation at lowest order of 
Planck mass could be expressed as
\begin{equation}
\omega^2 = k^2 \pm \xi \frac{k^3}{M_P}
\label{eq3}
\end{equation}
where $\omega$ and $k$ are the photon's energy and momentum, respectively,
while the $\xi$ is a dimensionless parameter and the $\pm$ signs stand for 
opposite helicity photons~\cite{Gambini:1998it}. 

Several theories are predicting or supporting the relation (\ref{eq3}). 
The Planck-scale quantum gravity modifies the Maxwell equations by 
adding extra terms proportional to the Planck length~\cite{Gleiser:2003fa}:
\begin{equation}
\frac{\partial{\vec{E}}}{\partial{t}} = \vec{\nabla}\times\vec{B}-2\xi l_{P} 
\vec{\nabla}^{2} \vec{B}
\label{eq1}
\end{equation}
\begin{equation}
\frac{\partial{\vec{B}}}{\partial{t}} = -\vec{\nabla}\times\vec{E}-2\xi l_{P} 
\vec{\nabla}^{2} \vec{E}
\label{eq2}
\end{equation}
which leads to a deformed energy-momentum or dispersion relation (\ref{eq3}).
In the above equations,  $\vec{E}$ and $\vec{B}$  describe the electromagnetic field.
More general expressions accounting for space anisotropy are derived 
in Ref.~\cite{Gubitosi:2010dj}. 
Using conventional definition $n=d\omega /dk$,
it is easy to verify that Eqs. (\ref{eq3})-(\ref{eq2})  
introduce a chiral vacuum with an energy dependent  birefringence
\begin{equation}
\Delta n = 3\cdot 10^{-19} \cdot \xi \cdot \omega [GeV]
\label{eq4}
\end{equation}
where the magnitude of $\xi$ defines the characteristic energies or distances where
quantum-gravity effects become sizable. 
In the simplest possible picture, this only happens at the Planck
scale, and hence $\xi=1$. However, the running of fundamental
constants with energy may require quantum gravity to become active a
few orders of magnitude below the Planck scale. The parameter $\xi$ is
there to account for such effects.

Another possible source of vacuum chirality is described by 
torsion gravity, an extension of the general relativity into the microscopic 
world to include particles' spins - for a review see~\cite{Hehl:1976kj}. 
In general, the spin gravity (space torsion) is considered to be weaker than the 
mass gravity (space curvature). However, near the Planck scale it may become detectable. 
Following Ref.~\cite{Prasanna:2009zz}, from the electromagnetic field Lagrangian 
\begin{equation}
{\mathcal L} =-\frac{1}{4}F_{\mu\nu}F^{\mu\nu}+qT^{\mu\nu\rho}
(\partial^\sigma F_{\mu\nu})F_{\rho\sigma}
\label{eq5}
\end{equation}
with a torsion tensor $T^{\mu\nu\rho}$ and free parameter $q$ one derives a dispersion 
relation quite similar to Eq.(\ref{eq3}): 
\begin{equation}
\omega^2 = k^2 \pm q S_0 k^3 
\label{eq6}
\end{equation}
where $S_0$ stands for a time component of the contorsion vector.

Myers and Pospelov~\cite{Myers:2003fd} derived the expression~(\ref{eq3}) 
within effective field theory with dimension 5 operators. 
A similar effect is calculated in Ref.~\cite{Dalvit:2000ay} exploring graviton 
interaction with an electromagnetic field in one-loop approximation.
In summary, chiral space is a universal feature of Planck-scale gravity, in
the sense that it is predicted by a large diversity of theories.

A nonbirefringent gravitational space is also possible
and has been  predicted within String theories using D-brane formalism. 
In Ref.~\cite{Ellis:2008gg}, a polarization independent refractivity 
\begin{equation}
 n-1=\zeta \frac{k}{M_P}
\label{eq7}
\end{equation}
is obtained for the space-time foam near the Planck length.
Here we use $\zeta$  instead of the $\xi$ to distinguish between the nonchiral and 
chiral space. In principle, both types may occur in the same vacuum at different scales  
$\zeta$ and $\xi$.
Both gravity induced effects, namely birefringence and
refractivity, share the common feature that their strength is growing
with the photon energy. This is in contrast to
the usual condensed matter or electromagnetic,
nontrivial vacua where the refraction effects are suppressed by powers of 
the energy~\cite{Dittrich:1998fy,Bombelli:2004tq}. 

{\it Current limits.---}
Experimental limits on space chirality are set by astrophysical observations  
exploring  birefringence induced depolarization of the linear light which comes from 
distant cosmological sources~\cite{Gleiser:2001rm}. The limits, however, are based on 
assumptions about the origin, spatial or temporal distribution
of the initial photons, and their possible interactions during the travel.  
Another critical assumption is a uniformly distributed birefringence 
over cosmological distances.
The most stringent limit $\xi < 2.4\cdot 10^{-15}$ is set 
by Ref.~\cite{Stecker:2011ps} based on photons with polarization $0.63\pm0.30$ in an energy 
range from 100 to 350~keV from GRB041219a~\cite{Glynn:2007}.
Sensitive particle-physics effects have been suggested to test
quantum gravity, mainly using threshold energies~\cite{Heyman:2003hs}.
Applying cosmic ray constraints on photon decay and vacuum Cherenkov 
radiation~\cite{Gharibyan:2003fe}, one arrives to $\zeta < 30$ and 
$\zeta < 300$ limits, respectively.

For the space refractivity, there are astrophysical observations interpreted~\cite{Ellis:2009yx} 
as $\zeta \sim 10$. This is derived from energy dependent time delay measurements of  photons 
from  distant sources. Similar to the results derived from polarized photons of
cosmological origin, strong assumptions have to be made on the source of these photons. 
In addition,  quoted astrophysical constraints are 
valid only for photon-virtual graviton loop interactions, since the photon path
is essentially free from gravitational fields.   

Terrestrial measurement could shed light on the quantum-gravity
promoted space chirality and refractivity including effects introduced by real, Earth 
gravitons. In the laboratory the Planck scale
can be accessed by exploring the extreme sensitivity of the high energy Compton 
scattering to the vacuum refraction as discussed in the following.

{\it Compton scattering affected by gravity.---}
Let us denote by $\omega_0$, $\omega$, $\theta_0$ and $\theta$ the energies and angles of the
incident and scattered photons relative to the initial electron direction.
Then, according to Ref.~\cite{Gharibyan:2003fe}, for the high energy Compton 
scattering in a vacuum with $n\approx 1$ (up to $\mathcal{O}[(n-1)^2]$ terms),
the energy-momentum conservation yields    
\begin{equation}
   n-1=\frac{\mathcal E}{2\gamma^2 ({\mathcal E}-\omega)}\Biggl(
   1+x+\theta^2\gamma^2-x\frac{\mathcal E}{\omega}\Biggr)
\label{eq8}
\end{equation}
where $\gamma$ and $\mathcal{E}$ are the Lorentz factor and energy of the initial
electron, \hbox{$x\equiv 4\gamma \omega_0\sin^2{(\theta_0/2)}/m$},
and $n$ is the refraction index for the direction $\theta$ and energy $\omega$.
This formula is more general than  Eq.(3) of Ref.~\cite{Gharibyan:2003fe}.
The difference is in a factor ${\cal E}/({\cal E}-\omega)$, because
in contrast to \cite{Gharibyan:2003fe} the 
final photon mass squared \hbox{$k_\mu^2 = \omega^2 (1-n^2)$} is not neglected
for this Letter.

Substituting $n-1$ in  Eq.(\ref{eq8}) by the gravitational refractivity from  Eq.(\ref{eq7}) 
we can estimate how the quantum gravity would change the scattered photons' maximal energy 
$\omega_{m}$ (Compton edge, at $\theta=0$).
The expected shift of the Compton edge is
\begin{equation}
\omega_m (n)-\omega_m (1)=\frac{32\gamma^6 \omega_0^2\sin^4{(\theta_0/2)}}{(1+x)^4} 
\frac{\zeta}{M_P}
\label{eq9}
\end{equation}
relative to the vacuum (n=1) kinematics.
At sufficiently high $\gamma$, the huge value of  $M_P$ is compensated, and the energy 
shift becomes detectable. Hence, this effect allows quantum-gravity 
induced space refractivity to be measured at accelerators by laser Compton scattering off 
high $\gamma$ electrons. 
For optical lasers and head-on collision the kinematic factor $x\approx 2\cdot 10^{-5}\gamma$
and the right-hand side of Eq.(\ref{eq9}) grows as $\gamma^6$ at $GeV$ energies 
slowing down to $\gamma^2$ growth above $TeV$ energies.  
\begin{figure}
\centering
\includegraphics[scale=0.45]{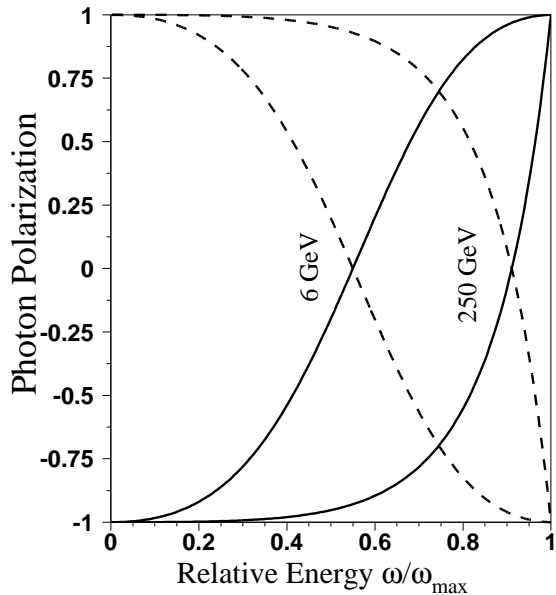}
\caption{\label{fig1}
Polarization of the Compton scattered photon on a 6 or 250~GeV electron as a function of 
the photon energy. The
solid and dotted lines correspond to the initial laser light helicity: +1 solid, -1 dotted.}
\end{figure}

In order to probe space birefringence, one needs to measure the
refractivity in Eq.(\ref{eq8}) for scattered photons of opposite helicity. 
This may be achieved by exploring circularly polarized initial laser beams and helicity 
conservation.  
The polarization of the secondary photons in the case of  scattering on
 unpolarized electrons is shown in Fig.\ref{fig1},  using formulas from 
Ref.~\cite{Lipps,mcmaster}.
At $\omega=\omega_{m}$ the polarization transfer is complete, such  
that the helicity of the Compton edge photons is fully defined by the laser light helicity. 
Consequently, in a birefringent vacuum the Compton edge energy is laser helicity dependent.
Evaluating Eq.(\ref{eq8}) for left and right helicity photons at $\theta=0$ yields
\begin{equation}
\Delta n = n_L (\omega_m^L) -  n_R(\omega_m^R) = \frac{(1+x)^2}{\gamma^2}A
\label{eq10}
\end{equation}
where $\omega_m^L$ and $\omega_m^R$ are the highest energies for the Compton  
opposite helicity photons and  \\
\hbox{$A=( \omega_m^L - \omega_m^R ) / ( \omega_m^L + \omega_m^R )$}
is an energy asymmetry.
 
Combining Eq.(\ref{eq10}) with the gravitational birefringence from  Eq.(\ref{eq4}), 
we arrive to
\begin{equation}
A = \frac{8\gamma^4 \omega_0 \sin^2{(\theta_0/2)}}{(1+x)^3} 
\frac{\xi}{M_P}
\label{eq11}
\end{equation}
which proves that for sufficiently high values of $\gamma$ the Planck-scale space  
birefringence generates a measurable asymmetry. 
For lasers with $\omega_0 \approx 3 eV$ and $\theta_0 = \pi$,  $\gamma$
dependence of  Eq.(\ref{eq11}) changes from $\gamma^4$ to $\gamma$ in the  
GeV to TeV range.

\begin{figure}
\centering
\includegraphics[scale=0.50]{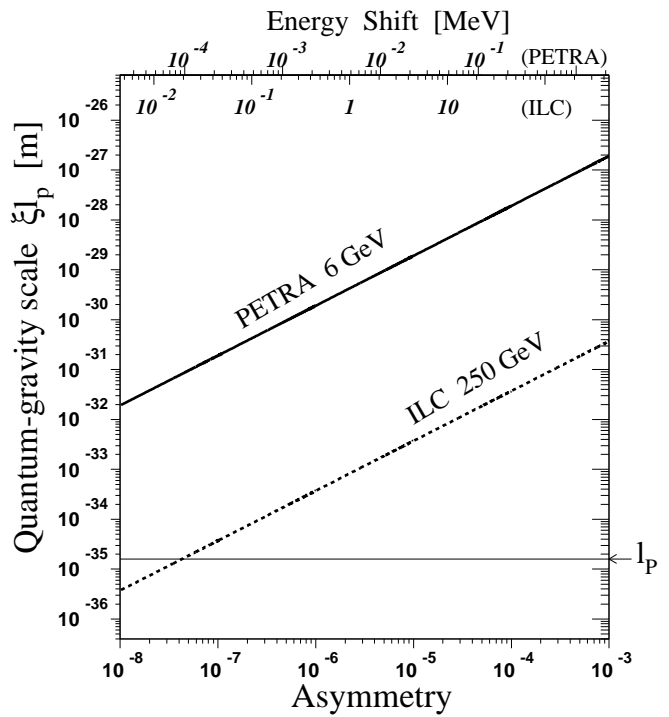}
\caption{\label{fig2}
Experimental reach of the accelerators for vacuum birefringence and refractivity.
Birefringence at the scale $\xi$ will produce a Compton edge asymmetry (lower scale)
while the refractivity produces absolute energy shifts (upper scale). The Planck 
length $l_P$ is shown by an arrow.}
\end{figure}

{\it Sensitivity of $e\gamma$ colliders.---}
To estimate sensitivity of the described methods we choose head-on collision
\hbox{($\theta_0=\pi$)} of an ordinary green laser \hbox{($\omega_0 = 2.41 eV$)}  with 
two different
energy electron beams, 6 \hbox{($x=0.22$)} and 250~GeV \hbox{($x=9.23$)}. The first 
energy is available at existent storage rings (e.g. PETRA-III~\cite{Balewski:2004iz}) 
while the higher energy may be accessible at the future International Linear 
Collider (ILC)~\cite{Djouadi:2007ik}. 
A graphical view of the experimental reach is presented in Fig.\ref{fig2}. 
By using Eq.(\ref{eq11}), the correspondence of the quantum-gravity scale 
 ($\xi l_P = \xi/M_P$) to the induced asymmetry is plotted. 
It is worth noting that
asymmetries as small as $10^{-7}$ have been detected with a sensitivity $10^{-8}$ 
at the SLAC 50~GeV experiments~\cite{Anthony:2003ub,Anthony:2005pm},   based
on beam helicity flips. 
Thus, similar accuracies seem reachable at PETRA or at the ILC which suggests
that the 6~GeV machine could test  space 
birefringence down to $10^{-31}m$ while the 250~GeV machine reaches the Planck 
length.  

 The space refractivity measurement is more difficult, since one needs to detect an absolute 
energy shift which is relatively small. For that the detector scale could be calibrated
by the bremsstrahlung edge~\cite{Gharibyan:2003fe}. An alternative could be the measurement of 
the scattered electron momentum using a spectrometer. Exploring lasers of different wavelengths 
may give even better sensitivity.
In either case the Compton edge detection is expected to be possible within an accuracy of 
\hbox{$\Delta \omega_m/\omega_m \sim 10^{-3}$}, thus probing  
space refractivity down to $10^{-27}$ and $10^{-31}m$ for the 6 and 250~GeV 
machines respectively.  

As a candidate for a quantum-gravity signature, consider high energy photon speed 
anomaly observations reported in 
Ref.~\cite{Gharibyan:2003fe}. Refractivities \hbox{$(1.69\pm 0.47)\times 10^{-11}$}
and \hbox{$(4.07\pm 0.05)\times 10^{-13}$} obtained with HERA 26.5~GeV~\cite{Barber:1992fc} 
and SLC 45.6~GeV~\cite{Shapiro:1993gd}
electron beams may be interpreted as quantum-gravity manifestation at 
\hbox{$(2.57\pm 0.71)\times 10^{-28}$} and \hbox{$(3.50\pm 0.04)\times 10^{-30}m$}
distances for the HERA and SLC measurements respectively.

In contrast to the astrophysical methods,  an accelerator Compton 
experiment is sensitive to the local properties of space at the laser-electron 
interaction point and along the scattered photon direction.
Hence, space isotropy tests are also possible as the accelerator rotates together 
with Earth. For any preferred direction, the measured  birefringence
is expected to change as the scattered photon beam sweeps a circle over the celestial sphere.
For a given direction ($\delta$,$\alpha$) of the photon beam and a possible anisotropy 
axis ($\delta_0$,$\alpha_0$), one expects  
\begin{equation*}
\Delta n = \Delta n_0 ( \cos{\delta} \cos{\delta_0}\cos{(\alpha-\alpha_0)}+\sin{\delta} \sin{\delta_0})
\label{eq12}
\end{equation*}
where $\Delta n_0$ is the maximal birefringence, along the 
declination $\delta_0$ and right ascension $\alpha_0$. 
Despite the  tight limits set by low energy high precision experiments on 
space anisotropy~\cite{Sudarsky:2002ue},  the accelerator isotropy test 
is a valuable and complementary test at high energies.
 
In conclusion, an accelerator based experiment is proposed to probe 
space birefringence and refractivity close to the Planck scale in photon
electron collisions.

I thank S.Schmitt for his careful and critical reading of the manuscript.

\end{document}